\documentclass[graybox]{svmult}

\usepackage{type1cm}        
\usepackage{makeidx}         
\usepackage{graphicx}        
\usepackage{multicol}        
\usepackage[bottom]{footmisc}

\usepackage{newtxtext}       %
\usepackage[varvw]{newtxmath}       

\usepackage{xspace} 
\newcommand{\etal}{et al}

\newcommand{\rrr}{\texttt{R3}\xspace}
\newcommand{\rr}{\texttt{R2}\xspace}
\newcommand{\xdraw}{\texttt{XDraw}\xspace}

\makeindex

\begin{document}

\title*{Visibility on Terrains}
\author{Laura Toma}
\institute{Laura Toma\at Bowdoin College, 255 Maine Street, Brunswick,
  ME, 04011, \email{ltoma@bowdoin.edu}}

%
%
\maketitle

\section*{Synonyms}

Viewshed; visible region; visibility map;  visualscape; viewscape

\section*{Definition}
\label{sec:def}

The simplest problem in terrain visibility is \emph{point-to-point
  visibility}: given a terrain model $T$, two arbitrary points points
$a, b$ on $T$, and possibly their heights/offset above the terrain,
determine whether $a$ and $b$ are visible to each other.  The line
segment $\bar{ab}$ between points $a,b$ is referred to as their
\emph{line-of-sight} (LOS); the points  are considered
visible if the interior of their LOS lies entirely above the terrain.

More generally, given a terrain model $T$, the location of an
arbitrary viewpoint $p$ on $T$, and possibly the height of $p$ above
the terrain, we want to find the \emph{viewshed}, or \emph{visible
  region} of $p$, which is the part of the terrain visible from $p$.

Viewsheds can be computed from several viewpoints at once, and are
referred to as \emph{cumulative viewsheds}: given a set of points on a
terrain, their cumulative viewshed is the part of the terrain that is
visible to at least one of the points, or, put differently, the union
of their viewsheds. 
Point-to-point visibility, viewsheds and cumulative viewsheds can be
computed on both grid and TIN terrains. On grid terrains specifically
the concept of \emph{total viewshed} gives a characterization of the visibility
of every pixel in the grid: the total viewshed is another grid of the
same size as the elevation grid, whose value at each pixel $u$ is the
size of the viewshed of $u$ (i.e. the number of grid points visible
from $v$).
For further visibility related concepts  see~\cite{floriani:vissurvey,ht:terrainsurvey2014}.

\section*{Historical Background}
\label{sec:backgr}

According to Inglis et al~\cite{inglis-2022} the term \emph{viewshed}
was introduced in 1967 by surveyor and landscape architect
C. Tandy~\cite{tandy-isovist}. A year later, in 1968, researchers at
the U.S. Forest Service developed a Fortran program called
ViewIt~\cite{amidon-viewit} to compute what part of a terrain was
visible from a given point.
Fast forward fifty years, with terrain data publicly available from
LiDAR technology and remote sensing at 1-2m resolution and high
accuracy, computing visibility on terrrains has been gaining in use
and spans a range of disciplines, from archaeology, landscape and
urban planning to military and defense applications.  Many GIS
software such as ESRI ArcGIS, QGIS, GRASS, provide functionality for
computing viewsheds.
%

Viewsheds have been studied on both TIN and grid terrains, although
the majority of applications and results use grids.
On grid terrains, the simplest form of viewshed is defined as a binary
grid where each cell is labeled as visible or not visible. Another
option is to determine,   for each visible point, its
vertical angle to visibility, or its \emph{visual exposure}; 
this tells not only what is visible, but whether visible points are
above or below the viewpoint and by how much (“how” visible they are).

From an algorithmic and complexity point of view, a binary
vs. angle-to-visibility viewshed are equivalent, and papers that aim
to improve the underlying algorithms do not distinguish between the
two.  For applications that focus on the quality of the viewshed and
generally how humans connect visually with their surrounding
environment, differentiating between binary viewsheds and the visual
exposure is important, and researchers have focused on further ways to
model factors that affect the quality of the visual space (e.g.  how
far the visible point is to the observer, where it is in their visual
field, what is next to it, lighting conditions, etc).
The term \emph{visualscape}, introduced by
Llobera~\cite{llobera-visualscapes}, aims to unify the possible ways
in which the visual space may be defined:
the viewshed, cumulative viewshed, and visual exposure are all types
of visualscapes.  Other types of visualscapes are the \emph{fuzzy
  viewsheds} or \emph{Higuchi viewsheds}, which assign lower weights
to points that are further away~\cite{higuchi1985visual}.

The concept of viewshed can be extended to terrain models that include
vegetation and built environment on top of the elevation layer, such
as available from LiDAR technology. When the terrain model includes
vegetation above the bare ground, researchers have looked at extending
the line-of-sight through tree canopies~\cite{llobera-vegetation}.
When the terrain model includes buildings and possibly vegetation on
top of the elevation, it is refered to as an \emph{urban scene}, and
the term used to represent the space visible to an observer is the
\emph{isovist}~\cite{benedikt-1979}.

Historically, obtaining viewsheds has been a computationally intensive
process.  Researchers have used sampling, lower resolution grids,  or
grids of smaller spatial extent in order to compute an approximate
cumulative or total viewshed in a reasonable amount of time.  At the
same time data is available at high resolution and accuracy, and it is
desirable that visibility algorithms be able to take advantage of it.  Many
researchers have worked to improve the theoretical and/or practical
efficiency of viewshed algorithms. Some results focus on obtaining
faster algorithms with theoretical guarantees; others proposed various
ways to speed up computation time by trading off accuracy of the
computed viewshed; recently, algorithms leverage high-performance
computing and GPUs.


\section*{Scientific Fundamentals}
\label{sec:fund}

\begin{figure}[b]
  \hfill 
  \includegraphics[width=5cm]{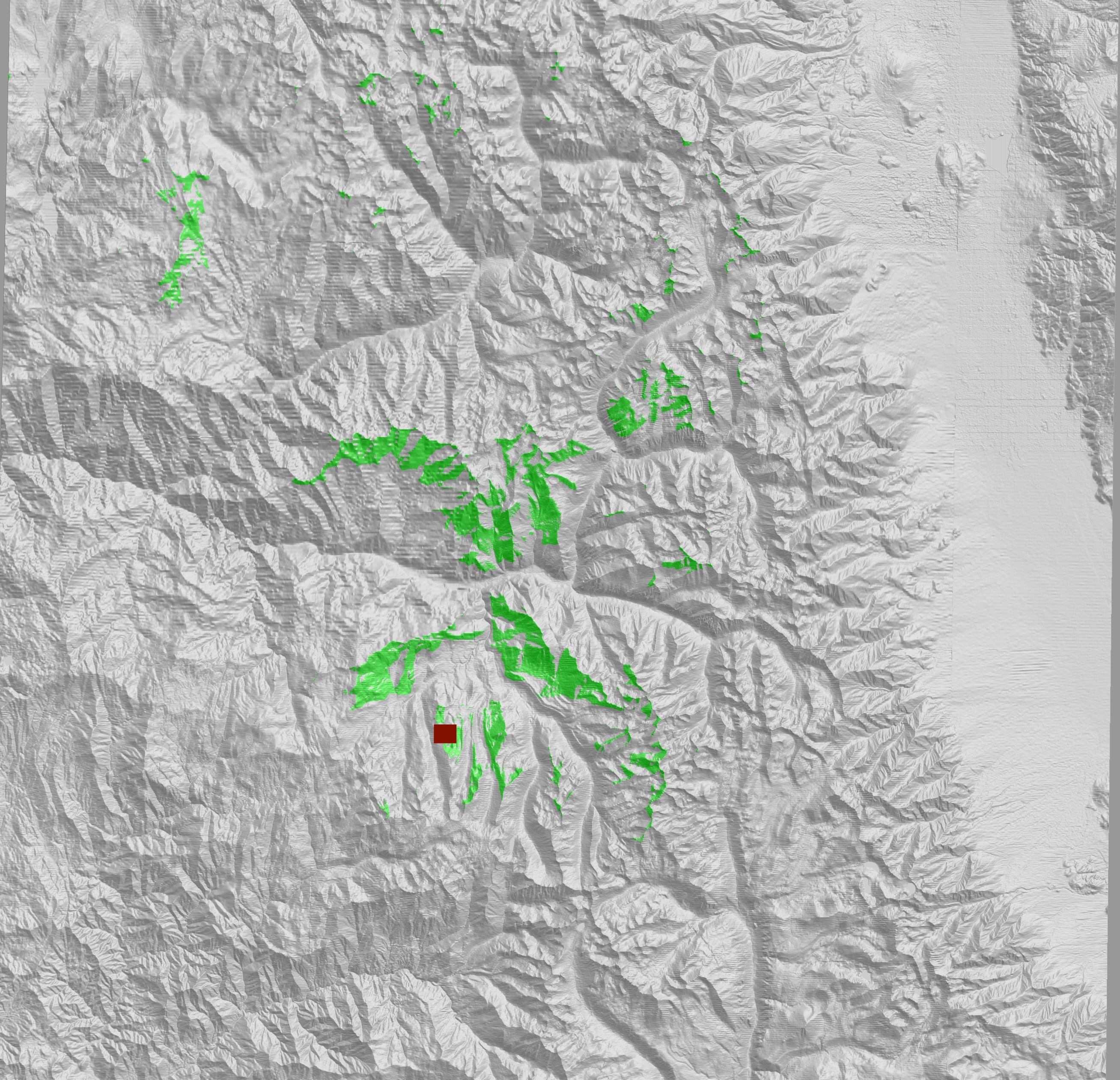} \hspace{1cm} 
  \includegraphics[height=4.8cm]{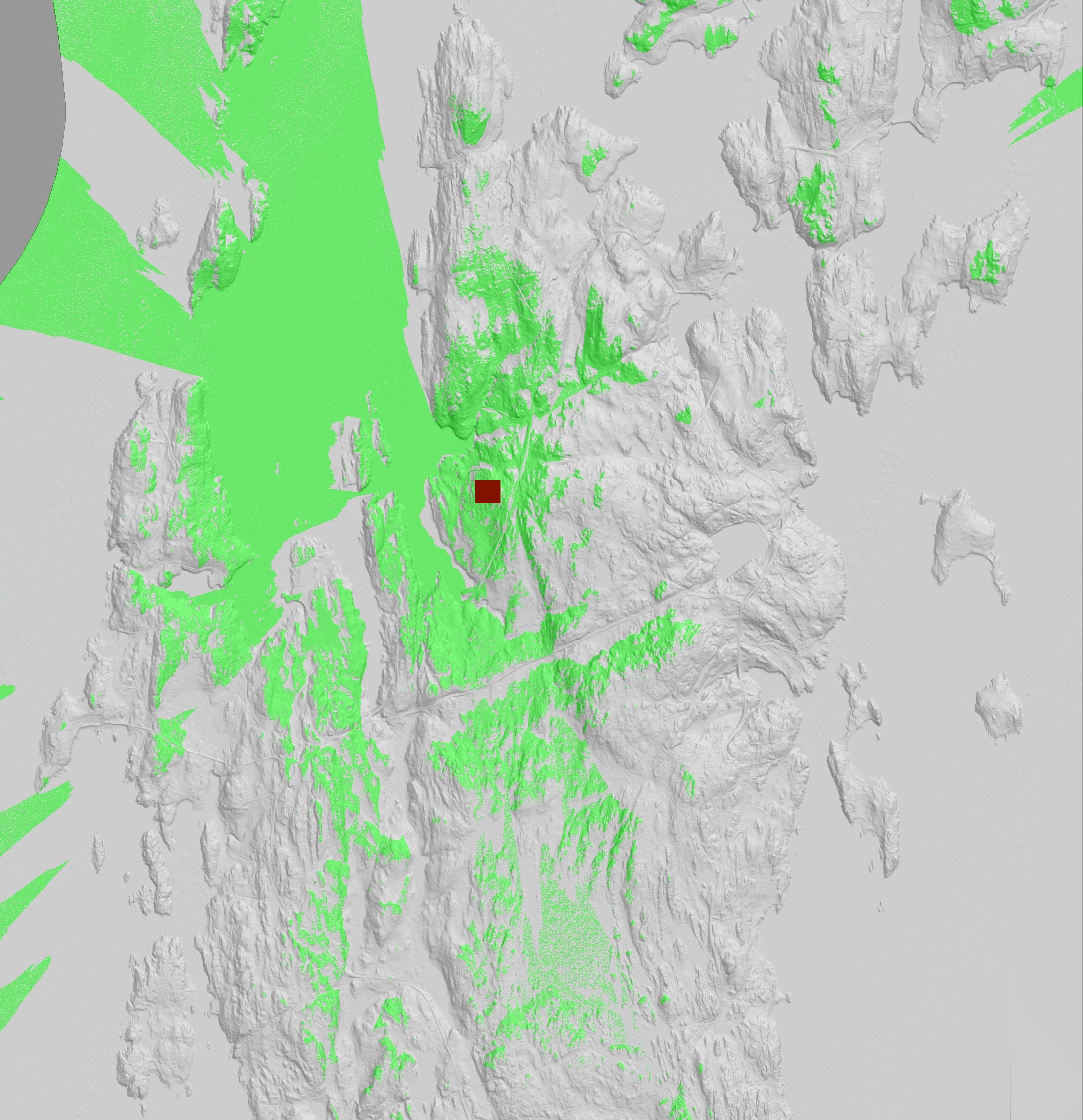}
  \hfill 
  \caption{ Two sample viewsheds in green with the viewpoints shown in red.  }
\label{fig:1}      
\end{figure}


A terrain is a surface in 2D (usually the bare ground elevation) with
the property that it intersects any vertical line in at most one
point.
A terrain model $T$ consists of a set of sample points $P$ and an
interpolation method that describes how to obtain the elevation (the surface value)
for points other than the sampled points $P$. On a grid terrain the
sample points form a regular grid, while on a TINs they have arbitary
position. We'll denote the number of sample points by $n$.
To decide whether two points $a$ and $b$ on $T$ are visible, one needs
to interpolate the elevation of $T$ along the projection of the LOS
$\bar{ab}$ on the horizontal plane. Thus, determining visibility
depends crucially on the underlying terrain model and specifically on
the interpolation method
\footnote{
In practice, terrain data comes as a discrete set of 3D sample
  points $P= \{p_1, p_2, ...p_n \} $, where each point is specified by
  its latitude and longitude and the measured elevation of $p_i$ which
  we denote by $z[p_i]$.  A terrain model consists of a set $P$ of
  sample points, together with an interpolation method that describes
  how to obtain the elevation values for points other than the sampled
  points $P$. More precisely, we need to define an elevation function
  $z$ that gives and an elevation value for any point $p$, that is,
  for any point $p$ that is not in $P$. This function $z$ should
  respect that data, that is, $z(p)$ should be equal to $z[p]$ for all
  $p \in P$.  }
\footnote{A grid terrain consists of a matrix of elevation values,
  representing elevation sampled from the terrain with a uniform grid;
  the $x$ and $y$ coordinates of the sample points are not stored
  explicitly as they can be inferred from the corner of the grid and
  resolution of the sampling.  A TIN terrain consists of an
  (irregular) sample of points, which are connected in a
  triangulation. Typically the triangulation forms a Delaunay
  triangulation when projected on the horizontal plane. Grids are the
  most commonly used model and the most widely available.
}.

Algorithms to compute viewsheds can be grouped based on the terrain
model (grid or TIN), and on the type of viewshed computed  (discrete
or continuous).  The \emph{discrete} viewshed of $p$ is the set of
points in $P$ that are visible from $p$; in other words, computing a
discrete viewshed means determining, for each sample point, whether it
is visible or not.
The \emph{continuous}, or \emph{geometric} viewshed
(or: visibility map, in computational geometry terms) of a viewpoint
$p$ is the union of all points on $T$ that are visible from $p$.
%
In order to be useful in practice, viewshed algorithms need to be fast
and accurate.  Efficiency is particularly important when computing
cumulative viewsheds and total viewsheds, or when computing viewsheds
on very large terrains with billions of points.  For a survey of
visibility results see~\cite{floriani:vissurvey,ht:terrainsurvey2014}.

%

 \paragraph{Visibility on TINs. }
A TIN terrain consists of a mesh of triangles and assumes linear
interpolation.  
The geometric viewshed on a TIN terrain is an arbitrary polygon whose
vertices are not necessarily vertices of the TIN and which can
intersect a triangle multiple times.
%
On a TIN of $n$ points, in the worst case there may be $\Theta(n)$
triangles such that the boundary of the visible part of each of them
consists of $\Theta(n)$ line segments, and thus a complete viewshed
may have size $\Theta(n^2)$~\cite{ColeSharir1989} (the  size of the viewshed is the number of vertices and edges on its boundary).
A quadratic complexity of the geometric viewshed would mean that,  in
practice,  computing it is unfeasible, especially on large
TINs. However, in practice, the complexity of the geometric
viewshed seems to be close to linear.
De Berg \etal~\cite{hh:smoothviewshed} were able to show that, under
some realistic  assumptions on the shape of the terrain,  the smoothed complexity of a viewshed is~$O(n)$ . 
%


To compute the geometric viewshed, the algorithm with the best
theoretical bound runs in
$O((n \alpha(n) + k)\lg n)$, where $k$ is the size of the viewshed,
and $\alpha()$ is the inverse Ackermann function~\cite{katz:hidden}.  When the size of
the viewshed is linear, $k=O(n)$, the running time becomes
$O(n  \alpha(n) \lg n )$.
%

A faster computation of the geometric viewshed is possible through
approximation.  Several such heuristics were described by Ben Moshe
\etal~\cite{moshe:approxvis, moshe:visibilitypreserving}. One idea is
to divide the area around the viewpoint into a number of slices with
each slice bounded by two rays from the viewpoint, and compute a
one-dimensional viewshed along each ray; if the visibility map of the
two rays bounding a slice are similar enough, the viewshed of the
whole slice is constructed by interpolation between the bounding rays;
otherwise the slice is subdivided into smaller ones.
Another idea is to simplify the terrain to reduce its size, and
compute visibility on the smaller, simplified terrain.  The
simplification process aims to preserve inter-point visibility (points
on the simplified terrain are visible if and only if they are visible
on the full-resolution terrain) and to this end it ensures that the
simplified terrain contains the important topographic features, such
as ridges and valleys. Both approaches were shown to give good results
in practice.

The geometric viewshed gives the most accurate represention of the
visible region, although it comes with the cost of potentially
quadratic size. One could argue that computing the full geometric
viewshed is not justified on a TIN that's not guaranteed to be a
perfectly accurate model of $T$. Instead, we can find the discrete
viewshed, i.e. the vertices of the TIN are visible from the viewpoint
$p$.  This can be phrased as a special instance of the ray shooting
problem and solved using the \emph{horizon tree}, by Cole and
Sharir~\cite{ColeSharir1989}. The horizon tree can be built in
$O(n \alpha(n)\lg n)$ time and has $O(n \alpha(n)\lg n)$ size. Given
the horizon tree, determining which of the $n$ vertices of the TIN are
visible from a viewpoint will take $O(n \lg n)$
time. 

\paragraph{Visibility on grids.}
Let $T$ be a terrain represented as a grid of $n$ points\footnote{For
  simplicity we assume that the grid is square and has size $\sqrt n$
  by $\sqrt n$.}.
A grid with linear interpolation corresponds to a triangular mesh, by
adding one of the diagonals in each grid cell.  Although geometric
viewsheds on grids can be computed with one of the TIN methods, they
have not been considered. The default viewshed on a grid terrain is
discrete and is represented as another grid, of the same size as $T$;
if a point $u$ is visible from viewpoint $p$, then $u$ is labeled as 1
in the viewshed grid\footnote{This can be easily extended to compute
  for e.g. a visibility angle instead of binary visibility}.

\begin{figure}[b]
  \includegraphics[width=3cm]{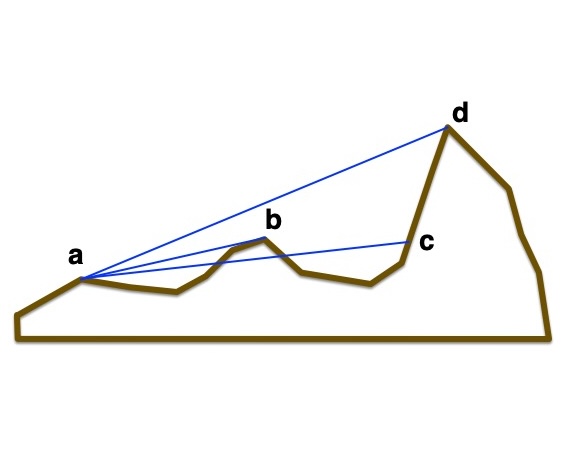}
  \includegraphics[width=3.5cm]{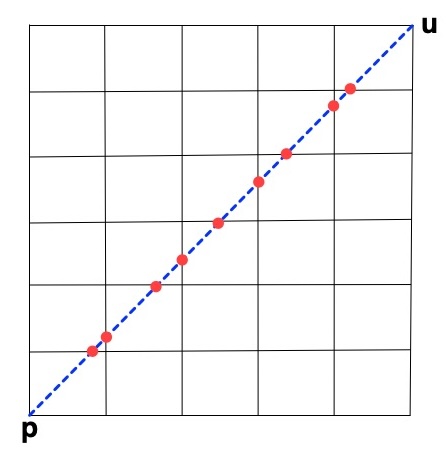}
  \includegraphics[width=5cm]{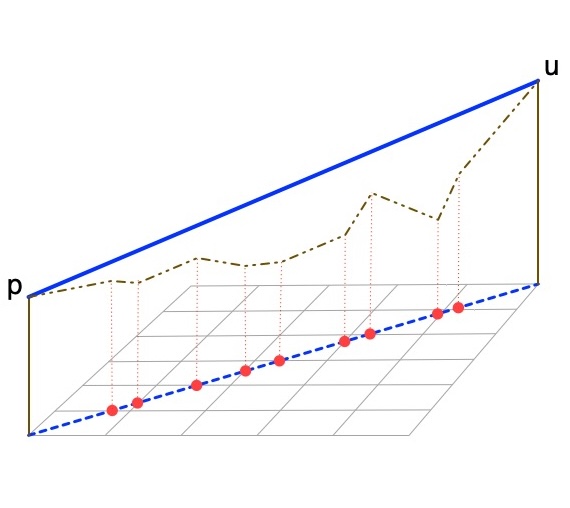}
  \caption{ (a) (a) A cross section of a terrain: points $b$ and $d$
    are visible from $a$, but not $c$.  (b) The projection of the LOS
    on the horizontal plane and the intersections with the grid lines;
    these are the points where the elevation of the terrain is
    interpolated to determine if $u$ is visible. (c) The viewshed in
    green with the viewpoint $p$ shown in red.  }
\label{fig:2}      
\end{figure}


The straightforward algorithm to find if a point $u$ is visible
from $p$ is to walk along the LOS $\bar{pu}$, compute the
intersections between the horizontal projection of $\bar{pu}$ and the
horizontal and vertical grid lines, and compute the elevation of the
terrain at these points with e.g. linear interpolation; point $u$ is
visible if all these elevations are below $\bar{pu}$; see
Figure~\ref{fig:2}.  We refer to this model as the \emph{gridlines
  model}.  Researchers noted that it corresponds to viewing the
terrain as consisting of grid cells bounded by the gridlines, with the
behaviour of the terrain inside the cells
unspecified~\cite{floriani:vissurvey}.
An alternative to the gridlines model is to model the terrain as a triangular mesh by
adding one of the diagonals in each grid cell.
In fact, Larsen~\cite{Larsen2015} showed that the the
gridlines model is  equivalent to adding, in each cell, the
diagonal that corresponds to the bottom of the tetrahedron spanned by
the four grid points.  
%
Either way, gridlines or triangles,
the visibility of a point $u$ can be determined in $O(\sqrt n)$ time.

The straightforward algorithm to compute the viewshed of $p$ is to
iterate over all grid points and check if they are visible from $p$.
This algorithm runs in $\Theta(n \sqrt n)$ time, is referred to as
\rrr, and was introduced by Franklin \etal~\cite{franklin:sdh94}; it
is considered to produce the ground truth viewshed
(e.g.~\cite{izraelevitz:vis}).

Two approaches are known that compute the same viewshed as \rrr, but
have a better time complexity~\cite{HaverkortTomaWei2013,
  PrescottToma2018}.  Both results traverse the gridline segments in
front-to-back order while maintaining horizons, based on Cole
\etal~\cite{ColeSharir1989} and De Floriani
\etal~\cite{floriani:horizon}. Put simply, the horizon $H_p$ of a set
of segments with respect to viewpoint $p$ is the upper rim of the
segments as they appear to a viewer at $p$. Haverkort et
al~\cite{HaverkortTomaWei2013} showed that on a grid terrain horizons
have complexity $O(n)$.
Their algorithm traverses the grid in layers around the viewpoint and
computes horizons in a divide-and-conquer fashion, running in overall
$O(n \lg n)$ time.  Prescott \etal~\cite{PrescottToma2018} extended
this horizon-based front-to-back computation to a two-step
process. Their observation is that viewsheds in practice are most of
the time very small, with large areas of the terrain invisible, and
one should be able to quickly filter out parts of the terrain that are
invisible. They proposed a two-step process: In the first step they
create a small, low-resolution grid to quickly filter out the parts of
the grid that are guaranteed to be invisible. In the second step they
use full resolution to find the points that are visible. This method
does not introduce any approximation and is an order of magnitude
faster than the previous algorithm from~\cite{HaverkortTomaWei2013} .


\emph{Approximate viewsheds:} Even with the faster algorithms
in~\cite{HaverkortTomaWei2013, PrescottToma2018}, computing the
discrete viewshed is too slow in practice when computing  for a large
number of viewpoints.  Many algorithms have been proposed that cut
down the computation and approximate the viewshed.
The algorithm known as \rr, examines the line-of-sight \emph{only} for
the $O(4 \sqrt n)$ grid points on the boundary of the grid; a grid
point that is not on the boundary of the grid is considered to be
visible if the nearest point of interesection between a grid line and
one of the examined lines-of-sight is determined to be visible.  \rr
runs in overall $O(n)$ time which makes it very appealing in
practice~\cite{franklin:sdh94}.  Researchers reported that it produces
a good approximation of \rrr that outweighs its loss in accuracy.
Another algorithm, called \xdraw, traverses the grid in concentric
layers~\cite{franklin:sdh94}.  When computing the visibility of the
points on a layer, it uses \emph{only} the grid points on the previous
layer.  More precisely, for a grid point $u$ in layer $i$, \xdraw
computes whether $u$ is visible by determining which are the two grid
points $q$ and $r$ in layer $i-1$ that are nearest to $pu$, and then
it estimates the maximum height above the horizon along $pu$ by
interpolating between the lines of sight to $q$ and $r$.  Thus, the
visibility of each point is determined in constant time per
point. \xdraw is faster than \rr, due to the simplicity of the
calculations, while being less accurate. A generalization of \xdraw
was described by Izraelevitz~\cite{izraelevitz:vis}.
Recently an improvement to \xdraw by reducing distortion was described
by Zhu at el~\cite{zhu-2019}.
Van Kreveld's radial sweep algorithm~\cite{kreveld:viewshed} runs in
$O(n \lg n)$ time but assumes the terrain is as a tessellation of flat
cells, where each cell is centered around a grid point and has the
same zenith across its span. The model is not ideal in practice as it
is not continuous, and some artifacts of the viewshed computed by this
algorithm were reported in~\cite{HaverkortTomaWei2013}.

\emph{Total viewsheds:} From a theoretical point of view, little is
known on computing the total viewshed faster than the straightforward
way, namely, compute the viewshed for every point in the grid. Any of the viewshed
algorithms mentioned above can be used and will result in
$O(n^2\lg n)$ and $O(n^2)$ running time, respectively.
One algorithm developed specifically for the total viewshed was
described by Tabik \etal~\cite{tabik:2015}. Their idea is to compute
the viewshed by diving the space around the point in a number of
sectors, each covering a small slice of azimuth angle; and process the
points in such an order so that sectors between consecutive points
overlap and the points can be reused. The resulting algorithm is
reported to be thousands of times faster than the viewshed tools in
ArcGIS and GRASS.


\paragraph{Parallel algorithms using multi-cores and GPUs} 

Recently research has focused on using high-performance computing to
reduce running times and enable computations for larger grids.
Parallel implementations of \rrr, \rr, and Van Kreveld's viewshed
algorithms were described
in~\cite{zhao2013:parallelr3,osterman2014:parallelr2,ferreira2014:parallelkreveld}.
%
%
%
The most impressive speedup was reported by Fernandez \etal~\cite{fernandez-2020-parallelTV} who extended their previous work on  the total viewshed~\cite{tabik:2015} to make it parallelizable. Based on an earlier idea of Stewart~\cite{stewart:1998}, the total viewshed is computed one sector at a time, for all points in the grid, and aggregated at the end. For a given sector, the idea is to re-arrange the grid so that instead of row-major order, it is stored so that any point in the grid finds the points in its sector consecutive in this order. This grid reordering, referred to as the \emph{skewed DEM}, improves the data locality and the memory accesses, and results in significant efficiency gains. Furthermore the whole computation can be easily parallelized and they reported a speedup of more than  800 times compared to the single-thread implementation.





\section*{Key Applications}
\label{sec:appl}


{\bf Archaeology.} Viewsheds can provide a lens into understanding
past cultures by reconstructing what might have been visible from
sites like fortresses, towers, and other cultural objects, which can
offer clues into power and control in past societies.  For
e.g. Malaperdas~\cite{malaperdas-2021} used viewsheds to reconstitute
the visual panorama of selected sites in the Kingdom of Pylos, in the
Peloponnese, Greece.  Mazuera et al~\cite{mazuera-2021} used
cumulative and total viewsheds to study the location of settlements in
the Sierra Nevada range in Colombia and found that the ability to see
the surrounding areas played a role in the location of these
settlements.
%

{\bf Visual exposure to greenery.} Viewsheds have been used to model
visual exposure to urban greenery and assess health and mental
benefits.  Cimburova et al~\cite{Cimburova2022}  described starting with a surface model grid and a
greenery grid for an urban area, generating  a number of sample
points from the greenery grid, computing  the viewpoint of each sample
point and combining  them into a cumulative viewshed to obtain a continuous map of
exposure to greenery. This tool was mplemented it as a tool in the open-source GIS
GRASS~\cite{GRASS-GIS} and is available for anyone to use. 

{\bf Touristic route visibility analysis. } Chamberlain et
al~\cite{chamberlain-2013} studied visibility along a popular tourist
highway route on Vancouver Island, and compared the standard
cumulative viewshed, where each point is labeled with the number of
times it is seen from the route, with their proposed \emph{visual
  magnitude (VM) index}, aimed at quantifying the quality of the view.
The VM analysis can identify areas in the landscape that are more
eye-catching along a route.

{\bf Visibility of water.} Qiang et al~\cite{qiang-2019}
described a method to compute ocean visibility for the island of
O'ahu, Hawaii. They created a sample of ocean points, computed a
viewshed for each point
and aggregated the results into a cumulative viewshed representing 
ocean visibility. They explored the effect of high-rise buildings on
ocean visibility, as well as the relation between ocean visibility and
a number of socio-economic and health variables such as income, age,
house prices and mental health.

{\bf Walkability analysis in cities.} Yang at
el~\cite{yang-2024-viewscape} used visibility to create walkability
maps and assess walkability of greenways in Detroit. 
Their idea was to create sample points along the greenways and compute
their viewshed/viewscapes using the R package
\emph{viewscape}~\cite{yang-2024-r}. Their integrated walkability
index at a point combines the size of the viewshed with other metrics
such as depth, relief, spatial openness, and proportion of
buildings/greenery in the viewshed.

{\bf Land conservation.} Anderson et al~\cite{anderson-2019}
developed a method to assign scores to land parcels bordering North
Carolina's Blue Ridge Parkway based on their visibility from the
parkway, where parcels with higher visibility were deemed higher
priority for conservation. To assess visibility they used a set of
sample points along the parkway and computed the cumulative viewshed
of these sample points.




\section*{Future Directions}
\label{sec:future}

\emph{A benchmark for evaluating viewshed algorithms.} A variety of
algorithms have been proposed to compute viewsheds, each one
representing a different trade-off between accuracy, speed in practice
and theoretical efficiency.
Developing a benchmark to assess the quality of the viewsheds computed
by the various existing methods is highly needed in order to focus and
accelerate future research.



\end{document}